\documentclass[12pt]{article}
\usepackage{graphicx}

\def\leaderfill{\leaders\hbox to 1em{\hss.\hss}\hfill}  

\begin{document}

\noindent {\footnotesize\it  ISSN 1063-7737,
 Astronomy Letters, 2007, Vol. 33, No. 11, pp. 720--728.
 \copyright Pleiades Publishing, Inc., 2007.
 \noindent Original Russian Text
 \copyright V.V. Bobylev, A.T.
Bajkova, 2007, published in Pis'ma v
Astronomicheski$\check{\imath}$ Zhurnal, 2007, Vol. 33, No. 11,
pp. 809--818.}

\noindent
\begin{tabular}{llllllllllllllllllllllllllllllllllllllllllllll}
& & & & & & & & & & & & & & & & & & & & & & & & & & & & & & & & & & & & &  \\
\hline \hline
\end{tabular}

\vskip 1.5cm

 \centerline {\large\bf Galactic Rotation Parameters from Data on Open Star
 Clusters}
 \bigskip
 \centerline {V.V. Bobylev, A.T. Bajkova, and S.V. Lebedeva}
 \bigskip
 \centerline {\small\it
Central (Pulkovo) Astronomical Observatory of RAS, St-Petersburg}
 \bigskip

{\bf Abstract--}Currently available data on the field of
velocities $V_r$, $V_l$, $V_b$ for open star clusters are used to
perform a kinematic analysis of various samples that differ by
heliocentric distance, age, and membership in individual
structures (the Orion, Carina--Sagittarius, and Perseus arms).
Based on 375 clusters located within 5 kpc of the Sun with ages up
to 1~Gyr, we have determined the Galactic rotation parameters
 $\omega_0  =-26.0\pm0.3 $ km s$^{-1}$ kpc$^{-1}$,
 $\omega'_0 = 4.18\pm0.17$ km s$^{-1}$ kpc$^{-2}$,
 $\omega''_0=-0.45\pm0.06$ km s$^{-1}$ kpc$^{-3}$,
 the system contraction parameter
 $K = -2.4\pm0.1$ km s$^{-1}$ kpc$^{-1}$,
 and the parameters of the kinematic center
 $R_0 =7.4\pm0.3$ kpc and $l_0 = 0\pm1^\circ$.
The Galactocentric distance $R_0$ in the model used has been found
to depend significantly on the sample age. Thus, for example, it
is $9.5\pm0.7$ kpc and $5.6\pm0.3$ kpc for the samples of young
($\leq50$ Myr) and old ($>50$ Myr) clusters, respectively. Our
study of the kinematics of young open star clusters in various
spiral arms has shown that the kinematic parameters are similar to
the parameters obtained from the entire sample for the
Carina-Sagittarius and Perseus arms and differ significantly from
them for the Orion arm. The contraction effect is shown to be
typical of star clusters with various ages. It is most pronounced
for clusters with a mean age of $\approx100$~Myr, with the
contraction velocity being $Kr = -4.3\pm1.0$ km s$^{-1}$.

\section*{INTRODUCTION}

The Galactic rotation parameters have been repeatedly determined
by many authors using objects belonging to various Galactic
structural components: from ionized and neutral hydrogen (Fich et
al. 1989; Merrifield 1992; Brand and Blitz 1993; Nikiforov 1999;
Avedisova 2005), from distant OB associations of stars (Dambis et
al. 2001; Mel'nik et al. 2001), and from open star clusters
(Zabolotskikh et al. 2002; Gerasimenko 2004; Popova and Loktin
2005b).

Open stars clusters (OSCs) are of great interest in studying the
kinematics of the Galaxy, since they are located in a wide solar
neighborhood and have reliable distance and age estimates.

Data only on the radial velocities $V_r$ (Mishurov et al. 1997;
Gerasimenko 2004; Popova and Loktin 2005b) are commonly used for a
kinematic analysis of OSCs, because the random errors in the
radial velocities are essentially distance-independent and, hence,
distant objects can be used. Note also that data only on the
radial velocities are available for distant hydrogen clouds (Fich
et al. 1989; Avedisova 2005).

When such catalogs as Hipparcos (1997) and Tycho-2 (Hog et al.
2000) appeared, it became possible to accurately determine the
mean proper motions of OSCs (Beshenov and Loktin 2004; Kharchenko
et al. 2005b). A number of authors used simultaneously two
observed velocity components, $V_r$ and $V_l$ (Mishurov and Zenina
1999; Dambis et al. 2001), or all three velocity components,
$V_r$, $V_l$, and $V_b$ (Zabolotskikh et al. 2002; Bobylev 2004),
to samples of stars within 3.4 kpc of the Sun. In this paper, we
also used all three observed velocity components.

One of the important problems is to determine the distance to the
center of Galactic rotation $R_0$. This parameter is estimated
indirectly from an analysis of the velocities of objects by
reconciling the adopted model with observational data. Only in
recent years has this parameter been estimated directly from
objects of the Galactic nucleus (McNamara 2000; Eisenhauer et al.
2003). A summary of the $R_0$ values determined in the last decade
by various methods can be found in Avedisova (2005) and an
overview of the $R_0$ estimation methods is given in Nikiforov
(2004).

The goal of this paper is to determine the Galactic rotation
parameters
 ($\omega_0$,
  $\omega'_0$,
  $\omega''_0$) and the parameters of the
kinematic center (Galactocentric distance $R_0$ and direction
$l_0$) from currently available data on the field of space
velocities $V_r$, $V_l$, and $V_b$, distances, and ages of OSCs
and to study the kinematic peculiarities of various samples
differing by heliocentric distance, age, and membership in
individual arms.

\section*{DATA}

At present, more than 1700 OSCs are known in the solar
neighborhood. Data on their proper motions, radial velocities, and
positions are needed for our purposes. A catalog that includes 652
OSCs (Kharchenko 2001; Kharchenko et al. 2005a, 2005b; Piskunov et
al. 2006) forms the basis for our work list. The advantage of this
catalog is a homogeneity and a high accuracy of the determination
of mean cluster proper motions achieved by using the ASCC-2.5 all
sky catalog (Kharchenko 2001) compiled from Hipparcos (1997),
Tycho-2 (Hog et al. 2000), and several other sources. The age
estimates obtained by comparison with isochrones are available for
the clusters of this catalog. The cluster distance estimates are
based on the results by Loktin and Beshenov (2003), who reconciled
the photometric estimates with the Hipparcos distance scale.We
took other data from the compilation by Dias et al. (2002) and the
WEBDA database (http://obsww.unige.ch/webda/).

For such open clusters as ASCC 16, ASCC 18, and Tr 10, we used the
mean radial velocities that we improved (Bobylev 2006) using the
OSACA catalog (Bobylev et al. 2006a).

As a result, we compiled a database on the proper motions, radial
velocities, and distances of 394 OSCs. They are located within
about 5 kpc of the Sun. Their ages do not exceed 1.5 Gyr. The
radial velocities with estimates of their random errors are
available for 270 ($\approx70\%$ of the sample) clusters. Note for
comparison, that Dias and Lepine (2005) used only 212 OSCs with
measured proper motions and radial velocities served as a source
for their kinematic analysis.

To identify clusters belonging to various structures (e.g., the
Orion, Perseus, and Carina-Sagittarius arms), we use a
probabilistic approach that we developed (Bobylev and Bajkova
2007). It is based on the approximation of the two-dimensional
($XY$) probability density function for the objects under
consideration by a set of Gaussians that represent the probability
density functions of individual features.

\section*{THE METHOD}

In this paper, we use a rectangular Galactic coordinate system
with the axes directed away from the observer toward the Galactic
center ($l=0^\circ$, $b=0^\circ$, the $X$ axis), along the
Galactic rotation ($l=90^\circ$, $b=0^\circ$, the $Y$ axis), and
toward the North Galactic Pole ($b=90^\circ$, the $Z$ axis).

The youngest disk objects that lie in the Galactic plane are
commonly used to determine the Galactic rotation parameters. As a
result, the observed velocity component $V_b$ is generally not
considered, while the component of solar motion $w_\odot$ is
assumed to be known (Dambis et al. 2001; Gerasimenko 2004). At the
same time, analysis of relatively distant objects (Zabolotskikh et
al. 2002; Bobylev 2004) showed that including the components
$w_\odot$ and $V_b$ in the model allows their values to be also
determined reliably. Therefore, we dwell on this more general
approach. In the special case where the clusters are distributed
in a ring (see the Carina-Sagittarius and Perseus sample below),
we used a fixed value of $w_\odot=7.2$ km s$^{-1}$ (Dehnen and
Binney 1998).

The method for determining the kinematic parameters used here
consists in minimizing the quadratic functional $F$,
$$
\displaylines{\hfill
 \min~~F=\sum_{i=1}^N w_r^i (V_r^i-\hat{V}_{r}^i)^2
 +\sum_{i=1}^N w_l^i (V_l^i-\hat{V}_{l}^i)^2+\sum_{i=1}^N w_b^i (V_b^i-\hat{V}_{b}^i)^2
\hfill\llap(1)
 }
 $$
under the following constraints derived from Bottlinger's formulas
(Ogorodnikov 1965) with the angular velocity of Galactic rotation
expanded in a series to terms of the second order of smallness in
$r/R_0$:
$$
\displaylines{\hfill
 V_r=-u_\odot\cos b\cos (l-l_0)-
 \hfill\llap(2)\cr\hfill
     -v_\odot\cos b\sin (l-l_0)
     -w_\odot\sin b-
 \hfill\cr\hfill
 -R_0(R-R_0)\sin (l-l_0)\cos b \omega'_0-
 \hfill\cr\hfill
 -0.5R_0 (R-R_0)^2 \sin (l-l_0)\cos b \omega''_0+
 \hfill\cr\hfill
 +r K \cos^2 b,\hfill
 \cr\hfill
 V_l= u_\odot\sin (l-l_0) - v_\odot\cos (l-l_0) -
 \hfill\llap(3)\cr\hfill
  -(R-R_0)(R_0\cos (l-l_0)-r\cos b) \omega'_0-
 \hfill\cr\hfill
  -(R-R_0)^2 (R_0\cos (l-l_0)-r\cos b)\times
 \hfill\cr\hfill
   \times 0.5\omega''_0 + r \omega_0 \cos b,\hfill\cr\hfill
 V_b=u_\odot\cos (l-l_0) \sin b +
 \hfill\llap(4)\cr\hfill
         +v_\odot\sin (l-l_0) \sin b
         -w_\odot\cos b +
  \hfill\cr\hfill
 +R_0(R-R_0)\sin (l-l_0)\sin b\omega'_0+
  \hfill\cr\hfill
 +0.5R_0(R-R_0)^2\sin (l-l_0)\sin b\omega''_0-
 \hfill\cr\hfill
 -r K \cos b\sin b,
\hfill
 }
$$
where $N$ is the number of clusters used; $i$ is the current
cluster number; $V_{r,l,b}$ are the cluster velocities to be
calculated, with $V_r$ being the radial velocity and
 $V_l= 4.74r\mu_l \cos b$ and $V_b= 4.74\mu_b$
 being the proper motion velocity components in the $l$ and
$b$ directions, respectively (the coefficient 4.74 is the quotient
of the number of kilometers in an astronomical unit by the number
of seconds in a tropical year);
 $\hat{V}_{r}^i, \hat{V}_{l}^i, \hat{V}_{b}^i$
 are the measured components of the velocity field;
 $w_r$,$w_l$,$w_b$
are the weights; $r$ is the heliocentric distance of the cluster
calculated via the photometric parallax determined by bringing the
cluster main sequence into coincidence with the corresponding
isochrone (Piskunov et al. 2006); the cluster proper motion
components $\mu_l \cos b$ and $\mu_b$ are in mas yr$^{-1}$, the
radial velocity $V_r$ is in km s$^{-1}$;
 $u_\odot$, $v_\odot$, $w_\odot$ are the cluster
centroid velocity components relative to the Sun; $R_0$ is the
distance from the Sun to the kinematic center of the system; $R$
is the distance from the cluster to the center of rotation; $l_0$
is the direction of the kinematic center; $R$, $R_0$, and $r$ are
in kpc. The quantity $\omega_0$ is the angular velocity of
rotation at distance $R_0$, the parameters $\omega'_0$ and
$\omega''_0$ are the first and second-order derivatives of the
angular velocity, respectively; $K$ is the Oort constant that
describes the expansion/contraction of the stellar system. The
distance $R$ can be calculated using the expression
$$
\displaylines{\hfill
 R^2=(r\cos b)^2-2R_0 r\cos b\cos (l-l_0)+R^2_0.\hfill\llap(5)
 }
$$
Note also that Eqs. (2)--(4) are written in such a way that the
direction of rotation from the $X$ axis to the $Y$ axis is
positive.

The weights in functional (1) are assigned in accordance with the
expressions (for simplification, the index $i$ was omitted)
$$
\displaylines{\hfill
 w_r=S_0/\sqrt {S_0^2+\sigma^2_{V_r}},
  \hfill\llap(6)\cr\hfill
 w_l=\beta^2 S_0/\sqrt {S_0^2+\sigma^2_{V_l}},
  \hfill\cr\hfill
 w_b=\gamma^2 S_0/\sqrt {S_0^2+\sigma^2_{V_b}},\hfill
 }
$$
where $S_0$ denotes the dispersion averaged over all observations,
which has the meaning of the ``cosmic'' dispersion that we take to
be 8 km s$^{-1}$;
 $\beta =\sigma_{V_r}/\sigma_{V_l}$ and
 $\gamma=\sigma_{V_r}/\sigma_{V_b}$
are the scale factors (in our case, $\beta=1$ and $\gamma=2$). The
errors in the velocities $V_l$ and $V_b$ can be calculated using
the formula
$$
\displaylines{\hfill
 \sigma_{(V_l,V_b)} = {4.74\over\pi}
 \sqrt{\mu^2_{l,b}\Biggl({\sigma_\pi\over\pi}\Biggr)^2+\sigma^2_{\mu_{l,b}}},\hfill
 }
$$
where $\sigma_\pi/\pi$ is taken to be 0.2, a typical
error in the photometric parallax.

In this paper, apart from the system of weights (6) described
above, we also use a variant of unit weights where $w_r=w_l=w_b=1$
for comparison.

The optimization problem (1)--(5) is solved for nine unknown
parameters $u_\odot$, $v_\odot$, $w_\odot$, $\omega_0$,
$\omega'_0$, $\omega''_0$, $K$, $R_0$, and $l_0$ by a
coordinate-wise descent method (the sought-for parameters are
taken as the coordinates). According to this method, the increment
$z_x$ for each unknown, which we arbitrarily denote by $x$, is
sought from a necessary condition for the existence of an extremum
of the functional $F(z_x)$:
$$
\displaylines{\hfill
 \frac{dF(z_x)}{dz_x}=0.\hfill\llap(7)
 }
$$
The functional $F(z_x)$ is obtained by substituting Eqs. (2)--(4)
for $V_r(z_x)$, $V_l(z_x)$, $V_b(z_x)$ into (1) using (5) in which
the variable $x$ is replaced by $x+z_x$.

Generally, Eq. (7) is a nonlinear equation for $z_x$. Its
numerical solution by the Newton method is an iterative process.
The value for $z_x$ at iteration $i$ is
$$
\displaylines{\hfill
 z_x^i=z_x^{i-1}
 -\Big({dF(z_x)}/{dz_x}\Big)
 /\Big({d^2 F(z_x)}/{dz_x^2}\Big)\Big|_{z_x=z_x^{i-1}}.\hfill
 }
$$
The solution for $z_x$ usually converges after 1--3 iterations. As
a result of one descent, the sought-for variable acquires a new
value, $x+z_x$. In our case, about 1000 descents in all nine
sought-for variables were required to obtain a solution with a
sufficiently high accuracy.

A sufficient condition for the existence of a global extremum is
that the Hessian matrix composed of elements
 $\{a_{i,j}\}=d^2F/dx_i dx_j$,
where $x_{i} (i=1,...9)$ denote the sought-for parameters, be
positively defined everywhere. We calculated the Hessian matrix in
a wide domain of parameters:
 $4\le u_\odot,v_\odot\le 16, 2\le w_\odot\le 11, -30\le\omega_0\le 0, -3\le \omega'_0\le 6, -0.6\le
 \omega''_0\le 0.3, -4\le K\le 5, 3\le R_0\le 15,-20\le l_0\le 10$.
Our analysis of the Hessian matrix for both weighting variants for
all of the cluster samples considered here showed it to be
positively defined. This suggests the existence of a global
minimum in this domain and, as a result, the uniqueness of the
solution. For unit weights, the Hessian matrix is also positively
defined far outside this domain, which could not be established in
the case of weighting according to rule (6). However, as will be
shown below, the adopted weighting allowed the accuracy of the
solutions obtained to be increased noticeably.

Interestingly, using weights of the form
 $w_{r, l, b}=1/\sigma^2_{V_{r, l, b}}$
provided no positive definiteness of the Hessian matrix in the
domain considered. It thus follows that several local extrema
exist in this case. Therefore, to find the correct solution, we
must know the initial approximation as accurately as possible. In
contrast, if a global extremum exists, then we can start from any
point of the domain. This, in particular, explains the
modification of the weights according to rule (6), which provided
a global minimum in a fairly wide domain of parameters.

We estimated the errors in the sought-for parameters by means of
Monte-Carlo simulations. The errors were estimated by performing
100 cycles of computations. At this number of cycles, the mean
values of the solutions virtually coincide with the solutions
obtained only from the input data, i.e., without any addition of
the measurement errors.

\section*{CONSTRAINTS}

In solving the optimization problem (1)--(5), we used the
following constraints on the data: (i) the magnitude of the
peculiar velocity $|V_{pec}|<90$ km s$^{-1}$; (ii) the cluster
height above the Galactic plane $|z|<500$~pc. The velocity
$V_{pec}$ is calculated with respect to the local standard of rest
 $u_\odot=10.0$ km s$^{-1}$,
 $v_\odot= 5.3$ km s$^{-1}$,
 $w_\odot= 7.2$ km s$^{-1}$
(Dehnen and Binney 1998). According to constraint (i), four
clusters were rejected. Constraint (ii) is important only for
clusters older than 50 Myr. According to this criterion, the seven
oldest clusters were rejected; all of the remaining clusters
proved to be younger than 1 Gyr. After applying an additional
constraint on the sample radius, $r<5$~kpc, 375 clusters remained
out the originally compiled list of 392 clusters, which served as
the material for a further analysis.

\section*{RESULTS}

Figure 1 presents the spatial distribution of OSCs in the Galactic
$XY$ plane. Since a kinematic analysis of OSC samples with age
separation is also the goal of this paper, we separately displayed
the distributions of OSCs younger and older than 50 Myr in Figs.
1a and 1b, respectively. We see from Fig. 1a that young clusters
clearly reproduce three segments of the spiral arms (Mel'nik et
al. 2001; Dias and L\'epine 2005; Popova and Loktin 2005a). The
concentration regions are marked in the figure by the ellipses:
the Perseus arm is on the left, the Orion arm is at the center,
and the Carina-Sagittarius arm is on the right. We clearly see
that the distribution of older clusters (Fig. 1b) is considerably
more uniform, more compact, and concentrated closer to the Sun.

Figure 2 shows the mean errors $\sigma_{V_r}$, $\sigma_{V_l}$, and
$\sigma_{V_b}$ as a function of the heliocentric distance. Note
that in solving the optimization problem (1)--(5), we take into
account the peculiarities of the error distribution using the
weighting procedure (6).

Table 1 presents the results of our kinematic analysis of OSC
samples with different sample radii ($r<5$ kpc and $r<2.5$ kpc)
and age separation (younger and older than 50 Myr) using the two
weighting methods. Column 1 gives the sample type and size; column
2 lists the centroid velocity components relative to the Sun;
columns 3, 4, and 5 list the rotation parameters; columns 6, 7,
and 8 give the values of the $K$-effect, $R_0$, and $l_0$. The
results obtained with unit weights and weights (6) are presented
in the upper and lower parts of the table, respectively. As can be
seen from Table 1, the random errors in virtually all parameters
decrease when the system of weights (6) is used. For both
weighting methods, the difference in $R_0$, which is significant
for the sample of old clusters, is largest. As can be seen from
Table 1, we obtain a discrepancy in $\omega''_0$ depending on the
constraint on the sample radius. Figure 3 shows how the parameter
$\omega''_0$ affects the Galactic rotation curve. To properly
compare the results and to construct the curves indicated in Fig.
3, we obtained two solutions at fixed $R_0=7.4$ kpc:
$$
\displaylines{\hfill
   \omega_0=-26.7\pm 0.3~\hbox{km s$^{-1}$ kpc$^{-1}$}, \hfill\llap(8)\cr\hfill
  \omega'_0= 4.27\pm0.11~\hbox{km s$^{-1}$ kpc$^{-2}$}, \hfill\cr\hfill
 \omega''_0=-1.05\pm0.06~\hbox{km s$^{-1}$ kpc$^{-3}$}, \hfill
}
$$
for the $r<2.5$ kpc sample and
$$
\displaylines{\hfill
   \omega_0=-26.0\pm 0.2~\hbox{km s$^{-1}$ kpc$^{-1}$}, \hfill\cr\hfill
  \omega'_0= 4.14\pm0.10~\hbox{km s$^{-1}$ kpc$^{-2}$}, \hfill\cr\hfill
 \omega''_0=-0.44\pm0.04~\hbox{km s$^{-1}$ kpc$^{-3}$}  \hfill
}
$$
for the $r<5$ kpc sample. As can be seen from Fig. 3, the
difference in the rotation curve due to the influence of the
second derivative becomes noticeable as one receded from $R_0$ to
distances exceeding $\approx1.5$ kpc.

The Oort constants
 $A=0.5 R_0 \omega'_0$ and
 $B= \omega_0+0.5 R_0 \omega'_0$ calculated from the data in the fifth row of Table 1 are
 $A= 15.4\pm0.6$ km s$^{-1}$ kpc$^{-1}$ and
 $B=-10.6\pm0.7$ km s$^{-1}$ kpc$^{-1}$ and
have no significant differences for all solutions of Table 1.

In Fig. 4, linear contraction velocity $Kr$ is plotted against
heliocentric distance $r$ and mean sample age $t$. To obtain these
characteristics, we divided the entire data set into four groups
in such a way that each group contained approximately the same
number of clusters. The dependences in Fig. 4 were derived at
fixed Galactocentric distance $R_0=7.4$ kpc. It follows from Fig.
4a that the contraction effect is traceable at all heliocentric
distances $r>1$ kpc and is, on average, $Kr = -3.2\pm1.0$ km
s$^{-1}$. As we see from Fig. 4b, the contraction effect is
typical of star clusters with various ages, but it is most
pronounced for clusters with a mean age $t\approx100$ Myr; the
contraction velocity in this case is $Kr =-4.3\pm1.0$ km s$^{-1}$.

To study the kinematic peculiarities of various structures, we
analyzed clusters younger than 50 Myr belonging to various arms
(Fig. 1a). To identify the arms, we calculated the probability
that each cluster belonged to one or another arm and then
attributed it to the arm the probability of belonging to which was
at a maximum. In this case, the probability densities of cluster
belonging to each arm were fitted by Gaussians whose bases in the
shape of ellipses are shown in Fig. 1b. The results of our
kinematic analysis in which all parameters were found
simultaneously are presented in the first four rows of Table 2.
The first and second two rows give, respectively, the results
obtained with unit weights and with weights (6).

For the Carina-Sagittarius and Perseus sample, which contains
mostly distant clusters, we also obtained a solution at fixed
 $w_\odot=7.2$ km s$^{-1}$ using weighting (6). This solution is given in the last
row of Table 2. It shows that a change in parameter $w_\odot$ by
$\sim2$ km s$^{-1}$ does not affect significantly the
determination of other model parameters.

\section*{DISCUSSION}
\subsection*{Parameters of the Kinematic Center}

As can be seen from Table 1, the values of $R_0$ obtained with
various weighting methods agree well, within the 1$\sigma$ error
limits. $R_0=7.4\pm0.3$ kpc that we found using clusters with a
mixed (in age) composition is in good agreement with the $R_0$
estimates made by various authors.

Thus, for example, in the opinion of Avedisova (2005), the most
probable value of $R_0$ lies within the range from 7.5 to 8.2 kpc.
By analyzing the $R_0$ determinations by various authors,
Nikiforov (2004) showed that the currently best value of $R_0$ is
$7.9\pm0.2$ kpc. Using 270 OSCs, Chen and Zhu (2007) obtained an
estimate of $R_0=7.95\pm0.62$ kpc.

Note that the Galactocentric distance $R_0=7.4\pm0.3$ kpc we found
provides more evidence for the short distance scale (Dambis et al.
2001) than the IAU recommendations (1985), $R_0=8.5$ kpc.

At the same time, as we see from Table. 1, $R_0$ depends
significantly on the sample age. Thus, for example, it is
$R_0=9.5\pm0.7$ kpc for clusters younger than 50 Myr. On the other
hand, a value of $R_0=7.1\pm0.6$ kpc, which is close to that found
from all 375 OSCs, was found from clusters younger than 50 Myr
that belong only to the Carina-Sagittarius and Perseus arms, but
not to the Orion arm (Table 2). Our separate study of the Orion
arm showed a significant difference almost in all kinematic
parameters (see Table 2).

Clusters older than 50 Myr are distributed fairly uniformly in
space (Fig. 1b); the influence of the Galactic spiral structure
and the Local system of stars in their motions is minor. One would
think that they should be well suited to a reliable determination
of $R_0$, but we found $R_0=5.6\pm0.3$ kpc for them (the lower
part of Table 1). Based on a kinematic analysis of the OSC radial
velocities from various catalogs and, in particular, the catalog
by Piskunov et al. (2006), Nikiforov and Kazakevich (2006) point
out a great uncertainty in $R_0$. In particular, they found
$R_0=6\pm0.7$ kpc from a sample of old
 ($\log t>8.8$) clusters, in good agreement with our result.

Analysis of the motions of the stars nearest the Sun showed that
the vertex deviation for the youngest stars reaches $30^\circ$
(Dehnen and Binney 1998). Based on the clusters the results of
whose analysis are presented in Table 1, we found no significant
deviation from the direction $l_0=0^\circ$. This means that the
rotation of the Galactic disk is nearly axisymmetric on large
scales.

\subsection*{The Galactic Rotation Curve}

The parameters of the Galactic rotation curve that we found in
solution (8) at fixed $R_0=7.4$ kpc (which, in our case, is best
suited to a proper comparison) are in good agreement with
 $\omega_0  =-27.5\pm1.4 $ km s$^{-1}$ kpc$^{-1}$,
 $\omega'_0 = 4.54\pm0.24$ km s$^{-1}$ kpc$^{-2}$,
 $\omega''_0=-1.09\pm0.19$ km s$^{-1}$ kpc$^{-3}$
determined from open clusters by Zabolotskikh et al. (2002) for
the short distance scale and for $R_0=7.5$ kpc. There is also good
agreement with the results of our analysis of distant OB stars
(Bobylev 2004):
 $\omega_0  =-28.0 \pm 0.6$ km s$^{-1}$ kpc$^{-1}$,
 $\omega'_0 =  4.17\pm0.14$ km s$^{-1}$ kpc$^{-2}$,
 $\omega''_0= -0.81\pm0.12$ km s$^{-1}$ kpc$^{-3}$
 (for $R_0=7.1$ kpc).

We found the parameter $\omega''_0$ to decrease in absolute value
with increasing sample radius. Far from the Sun, the Galactic
rotation curve constructed from our data (Fig. 3) disagrees with
the rotation curve constructed, for example, from the molecular
gas (Avedisova 2005), which is flat in a wide $R$ range, from 2
kpc to 15 kpc. Therefore, using a larger number of terms in the
expansion of $\omega_0$ in terms of $r/R_0$ is of considerable
interest in establishing the shape of the Galactic rotation curve
from OSCs far from the Sun.

\subsection*{The Contraction Velocity}

It is interesting to compare the contraction parameter
 $K=-2.4\pm0.1$ km s$^{-1}$ kpc$^{-1}$ that we found from all
clusters with the results by Torra et al. (2000). Based on the
Ogorodnikov-Milne linear model, these authors found
 $K=-2.0\pm0.4$  km s$^{-1}$ kpc$^{-1}$ for OB stars of all ages in the range of
distances 100--2000 pc and
 $K=-5.1\pm1.5$ km s$^{-1}$ kpc$^{-1}$ for OB
stars with ages $t>60$ Myr in the range of distances 100--2000 pc.
Fern\'andez et al. (2001) and Bobylev et al. (2006b) showed that
including the spiral structure did not eliminate the $K$-effect.
Based on our results (Fig. 4), we can assume that the observed
contraction velocity reflects the filling rate of the interarm
space (Marochnik and Suchkov 1984), but a detailed study of this
effect is outside the scope of our paper.

\subsection*{Kinematic Peculiarities of the Arms}

In this paper, we applied no corrections for the influence of the
spiral structure to the observed velocities. However, the
influence of the spiral structure was studied in our previous
paper (Bobylev et al. 2006b), where we showed that it has a major
effect on the centroid velocity components relative to the Sun.
This can explain the significant difference in the components
 $\Delta u_\odot\approx2$ km s$^{-1}$ and
 $\Delta v_\odot\approx5$ km s$^{-1}$ found from
various arms.

Analysis of the kinematic parameters for young clusters belonging
to the Orion arm indicates that such parameters as
 $\omega_0=-31.7\pm0.6$ km s$^{-1}$ kpc$^{-1}$ and
 $l_0 =-3\pm1^\circ$ differ significantly from the
Galactic rotation parameters found from all clusters. This shows
that the Orion arm has an additional rotation with an angular
velocity of $\approx5$ km s$^{-1}$ kpc$^{-1}$ around a center that
may not be associated with the Galactic center.

\section*{CONCLUSIONS}

We compiled a database on 375 OSCs from various present-day
catalogs and published sources. The clusters considered are
located with $<5$ kpc of the Sun. Their ages do not exceed 1 Gyr.

The following Galactic rotation parameters were determined from
data on the field of velocities $V_r$, $V_l$, $V_b$ for these
OSCs:
 $\omega_0  =-26.0\pm0.3 $ km s$^{-1}$ kpc$^{-1}$,
 $\omega'_0 = 4.18\pm0.17$ km s$^{-1}$ kpc$^{-2}$,
 $\omega''_0=-0.45\pm0.06$ km s$^{-1}$ kpc$^{-3}$,
 the system contraction parameter
 $K = -2.4\pm0.1$ km s$^{-1}$ kpc$^{-1}$,
 and the parameters of the center of rotation
 $R_0=7.4\pm0.3$ kpc and $l_0=0^\circ$.

In addition, we performed a kinematic analysis of various OSC
samples differing by heliocentric distance, age, and membership in
individual arms. The value of $R_0$ was found to depend on the
sample age. Thus, for example, it is $9.5\pm0.7$ kpc for clusters
younger than 50 Myr and $5.6\pm0.3$ kpc for clusters older than 50
Myr. Our study of the kinematics of young OSCs in various arms
showed that the derived kinematic parameters are similar to the
parameters obtained from the entire OSC sample for the
 Carina-Sagittarius arms and differ significantly from them for the Orion
arm. The contraction effect was shown to be typical of star
clusters with various ages and to be most pronounced for clusters
with a mean age of $\approx100$ Myr; the linear contraction
velocity in this case is $Kr=-4.3\pm1.0$ km s$^{-1}$.

\section*{ACKNOWLEDGMENTS}

We wish to thank I.I. Nikiforov for a discussion of our results
and the referees for helpful remarks that contributed
significantly to an improvement of the paper. This work was
supported by the Russian Foundation for Basic Research (project
no. 05-02-17047).

\section*{REFERENCES}

{\small

~~~~~1. V.S. Avedisova, Astron. Zh. 82, 488 (2005) [Astron. Rep.
49, 435 (2005)].

2. G.V. Beshenov and A.V. Loktin, Astron. Astrophys. Trans. 23,
103 (2004).

3. V.V. Bobylev, Pis'ma Astron. Zh. 30, 185 (2004) [Astron. Lett.
30, 159 (2004)].

4. V.V. Bobylev, Pis'ma Astron. Zh. 32, 906 (2006) [Astron. Lett.
32, 816 (2006)].

5. V.V. Bobylev, G.A. Goncharov, and A.T. Bajkova, Astron. Zh. 83,
821 (2006a) [Astron. Rep. 50, 733 (2006a)].

6. V.V. Bobylev, A.T. Bajkova, and G.A. Gontcharov, Astron.
Astrophys. Trans. 25, 143 (2006b).

7. V.V. Bobylev and A.T. Bajkova, Pis'ma Astron. Zh. 33, 643
(2007) [Astron. Lett. 33, 571 (2007)].

8. J. Brand and L. Blitz, Astron. Astrophys. 275, 67 (1993).

9. M. Chen and Z. Zhu, Chin. J. Astron. Astrophys. 7, 120 (2007).

10. A.K. Dambis, A.M. Melnik, and A.S. Rastorguev, Pis'ma Astron.
Zh. 27, 68 (2001) [Astron. Lett. 27, 58 (2001)].

11. W. Dehnen and J.J. Binney, Mon. Not. R. Astron. Soc. 298, 387
(1998).

12. W.S. Dias and J. R.D. L\'epine, Astrophys. J. 629, 825 (2005).

13. W.S. Dias, B.S. Alessi, A. Moitinho, et al., Astron.
Astrophys. 389, 971 (2002).

14. F. Eisenhauer, R. Schodel, R. Genzel, et al., Astrophys. J.
597, L121 (2003).

15. D. Fern\'andez, F. Figueras, and J. Torra, Astron. Astrophys.
372, 833 (2001).

16. M. Fich, L. Blitz, and A.A. Stark, Astrophys. J. 342, 272
(1989).

17. T.P. Gerasimenko, Astron. Zh. 81, 124 (2004) [Astron. Rep. 48,
103 (2004)].

18. The HIPPARCOS and Tycho Catalogues, ES ASP- 1200 (1997).

19. E. Hog, C. Fabricius, V.V. Makarov, et al., Astron. Astrophys.
355, L27 (2000).

20. N.V. Kharchenko, Kinematika Fiz. Nebesnykh Tel 17, 409 (2001).

21. N.V. Kharchenko, A.E. Piskunov, S. R\"{o}ser, et al., Astron.
Astrophys. 438, 1163 (2005a).

22. N.V. Kharchenko, A.E. Piskunov, S. R\"{o}ser, et al., Astron.
Astrophys. 440, 403 (2005b).

23. A.V. Loktin and G.V. Beshenov, Astron. Zh. 80, 8 (2003)
[Astron. Rep. 47, 6 (2003)].

24. L.S. Marochnik and A.A. Suchkov, Galaxies (Fizmatgiz, Moscow,
1984) [in Russian].

25. D.H. McNamara, J.B. Madsen, J. Barnes, et al., Publ. Astron.
Soc. Pacif. 112, 202 (2000).

26. A.M. Mel'nik, A.K. Dambis, and A.S. Rastorguev, Pis'ma Astron.
Zh. 27, 611 (2001) [Astron. Lett. 27, 521 (2001)].

27. M.R. Merrifield, Astron. J. 103, 1552 (1992).

28. Yu.N. Mishurov and I.A. Zenina, Astron. Astrophys. 341, 81
(1999).

29. Yu.N. Mishurov, I.A. Zenina, A.K. Dambis, et al., Astron.
Astrophys. 323, 775 (1997).

30. I.I. Nikiforov, Astron. Zh. 76, 403 (1999) [Astron. Rep. 43,
345 (1999)].

31. I.I. Nikiforov, Astron. Soc. Pac. Conf. Ser. 316, 199 (2004).

32. I.I. Nikiforov and E.E. Kazakevich, Astron. Astrophys. Trans.
25, 189 (2006).

33. K.F. Ogorodnikov, Dynamics of Stellar Systems (Fizmatgiz,
Moscow, 1958; Pergamon, Oxford, 1965).

34. A.E. Piskunov, N.V. Kharchenko, S. R\"{o}ser, et al., Astron.
Astrophys. 445, 545 (2006).

35. M.E. Popova and A.V. Loktin, Pis'ma Astron. Zh. 31, 190
(2005a) [Astron. Lett. 31, 171 (2005a)].

36. M.E. Popova and A.V. Loktin, Pis'ma Astron. Zh. 31, 743
(2005b) [Astron. Lett. 31, 663 (2005b)].

37. J. Torra, D. Fern\'andez, and F. Figueras, Astron. Astrophys.
359, 82 (2000).

38. M.V. Zabolotskikh, A.S. Rastorguev, and A.K. Dambis, Pis'ma
Astron. Zh. 28, 516 (2002) [Astron. Lett. 28, 454 (2002)]. }

\bigskip
Translated by V. Astakhov

{
\begin{table}[t]                                                
 \small
\caption[]{\small\baselineskip=1.0ex\protect
 Kinematic parameters of Galactic rotation
 }
\begin{center}
\begin{tabular}{|c|r|c|c|c|c|c|c|}\hline
 &&&&&&&\\
 Sample
 & $u_\odot, v_\odot, w_\odot$,
 & $\omega_0$,
 & $\omega'_0$,
 & $\omega''_0$,
 & $K$,
 & $R_0$,
 & $l_0$,\\
 & km/s~~~ & km/s/kpc & km/s/kpc$^2$ & km/s/kpc$^3$ & km/s/kpc & kpc & deg. \\\hline
 $r<5$    & $ 9.5_{(0.4)}$~~~ &                 &                 &                  &                &               &            \\
 kpc      & $10.9_{(0.3)}$~~~ & $-26.7_{(0.4)}$ & $3.91_{(0.21)}$ & $-0.35_{(0.08)}$ & $-2.3_{(0.1)}$ & $8.0_{(0.4)}$ & $ 0_{(1)}$ \\
 (N=375)  & $ 9.1_{(0.2)}$~~~ &                 &                 &                  &                &               &            \\\hline
 $r<2.5$  & $ 9.2_{(0.3)}$~~~ &                 &                 &                  &                &               &            \\
 kpc      & $11.9_{(0.3)}$~~~ & $-27.2_{(0.5)}$ & $3.72_{(0.32)}$ & $-0.70_{(0.13)}$ & $-2.6_{(0.2)}$ & $8.5_{(0.7)}$ & $-2_{(1)}$ \\
 (N=340)  & $ 8.4_{(0.2)}$~~~ &                 &                 &                  &                &               &            \\\hline
$t\leq50$ & $ 9.6_{(0.4)}$~~~ &                 &                 &                  &                &               &            \\
 Myr      & $11.7_{(0.4)}$~~~ & $-27.4_{(0.5)}$ & $3.21_{(0.25)}$ & $-0.14_{(0.07)}$ & $-1.4_{(0.2)}$ & $9.8_{(0.7)}$ & $ 0_{(1)}$ \\
 (N=196)  & $ 9.4_{(0.3)}$~~~ &                 &                 &                  &                &               &            \\\hline 
 $t>50$   & $ 9.7_{(0.4)}$~~~ &                 &                 &                  &                &               &            \\
 Myr      & $10.2_{(0.4)}$~~~ & $-25.3_{(0.9)}$ & $4.59_{(0.60)}$ & $-0.59_{(0.19)}$ & $-3.5_{(0.3)}$ & $6.7_{(0.8)}$ & $~0_{(1)}$ \\
 (N=178)  & $ 8.8_{(0.4)}$~~~ &                 &                 &                  &                &               &            \\
\hline 
\hline 
 $r<5$    & $ 9.2_{(0.2)}$~~~ &                 &                 &                  &                &               &            \\
 kpc      & $10.7_{(0.2)}$~~~ & $-26.0_{(0.3)}$ & $4.18_{(0.17)}$ & $-0.45_{(0.06)}$ & $-2.4_{(0.1)}$ & $7.4_{(0.3)}$ & $ 0_{(1)}$ \\
 (N=375)  & $ 8.2_{(0.2)}$~~~ &                 &                 &                  &                &               &            \\\hline
 $r<2.5$  & $ 8.8_{(0.2)}$~~~ &                 &                 &                  &                &               &            \\
 kpc      & $12.0_{(0.2)}$~~~ & $-26.7_{(0.3)}$ & $3.45_{(0.26)}$ & $-0.69_{(0.11)}$ & $-2.4_{(0.1)}$ & $9.0_{(0.7)}$ & $-2_{(1)}$ \\
  (N=340) & $ 8.0_{(0.1)}$~~~ &                 &                 &                  &                &               &            \\\hline
$t\leq50$ & $ 9.2_{(0.3)}$~~~ &                 &                 &                  &                &               &            \\
 Myr      & $11.5_{(0.3)}$~~~ & $-26.7_{(0.4)}$ & $3.24_{(0.22)}$ & $-0.19_{(0.07)}$ & $-1.4_{(0.2)}$ & $9.5_{(0.7)}$ & $ 0_{(1)}$ \\
 (N=196)  & $ 8.6_{(0.2)}$~~~ &                 &                 &                  &                &               &            \\\hline 
 $t>50$   & $ 9.4_{(0.3)}$~~~ &                 &                 &                  &                &               &            \\
 Myr      & $10.0_{(0.2)}$~~~ & $-25.5_{(0.5)}$ & $5.57_{(0.37)}$ & $-0.89_{(0.15)}$ & $-3.5_{(0.2)}$ & $5.6_{(0.3)}$ & $-1_{(1)}$ \\
 (N=178)  & $ 7.8_{(0.2)}$~~~ &                 &                 &                  &                &               &            \\
\hline
\end{tabular}
\end{center}
\end{table}
}
{
\begin{table}[t]                                                
 \small
\caption[]{\small\baselineskip=1.0ex\protect
 Kinematic parameters of young ($<50$ Myr) OSCs in the arms
 }
\begin{center}
\begin{tabular}{|c|r|c|c|c|c|c|c|}\hline
 &&&&&&&\\
 Arm
 & $u_\odot, v_\odot, w_\odot$,
 & $\omega_0$,
 & $\omega'_0$,
 & $\omega''_0$,
 & $K$,
 & $R_0$,
 & $l_0$,\\
 & km/s~~~ & km/s/kpc & km/s/kpc$^2$ & km/s/kpc$^3$ & km/s/kpc & kpc & deg. \\\hline
 Car-Sag     & $ 9.0_{(0.7)}$~~~ &                 &                 &                  &                &               &            \\
 and Perseus & $ 9.6_{(1.4)}$~~~ & $-27.3_{(0.6)}$ & $4.04_{(0.41)}$ & $-0.37_{(0.16)}$ & $-1.8_{(0.4)}$ & $8.3_{(1.0)}$ & $~1_{(1)}$ \\
 (N=92)      & $ 9.5_{(0.4)}$~~~ &                 &                 &                  &                &               &            \\\hline
 Orion       & $ 8.7_{(1.0)}$~~~ &                 &                 &                  &                &               &            \\
 (N=93)      & $14.2_{(0.4)}$~~~ & $-31.6_{(1.0)}$ & $4.46_{(0.84)}$ & $~0.23_{(0.38)}$ & $-0.5_{(0.6)}$ & $5.4_{(0.9)}$ & $-3_{(2)}$ \\
             & $ 9.5_{(0.3)}$~~~ &                 &                 &                  &                &               &            \\
 \hline\hline
 Car-Sag     & $ 6.5_{(0.5)}$~~~ &                 &                 &                  &                &               &            \\
 and Perseus & $ 8.7_{(1.1)}$~~~ & $-25.8_{(0.5)}$ & $4.52_{(0.37)}$ & $-0.70_{(0.17)}$ & $-2.6_{(0.4)}$ & $7.1_{(0.6)}$ & $ 1_{(1)}$ \\
 (N=92)      & $ 9.1_{(0.4)}$~~~ &                 &                 &                  &                &               &            \\\hline
 Orion       & $ 8.7_{(0.5)}$~~~ &                 &                 &                  &                &               &            \\
 (N=93)      & $14.0_{(0.4)}$~~~ & $-31.7_{(0.6)}$ & $5.14_{(0.65)}$ & $-0.20_{(0.34)}$ & $-0.5_{(0.4)}$ & $4.9_{(0.5)}$ & $-3_{(1)}$ \\
             & $ 8.3_{(0.3)}$~~~ &                 &                 &                  &                &               &            \\
\hline\hline
 Car-Sag     & $ 6.5_{(0.6)}$~~~ &                 &                 &                  &                &               &            \\
 and Perseus & $ 8.7_{(1.0)}$~~~ & $-25.9_{(0.5)}$ & $4.55_{(0.38)}$ & $-0.69_{(0.17)}$ & $-2.6_{(0.4)}$ & $7.1_{(0.7)}$ & $ 1_{(1)}$ \\
  (N=92)     &   --- ~~~~~  &                 &                 &                  &                &               &            \\
\hline
\end{tabular}
\end{center}
\end{table}
}

\newpage
\begin{figure}[t]
{
\begin{center}
  \includegraphics[width=160mm]{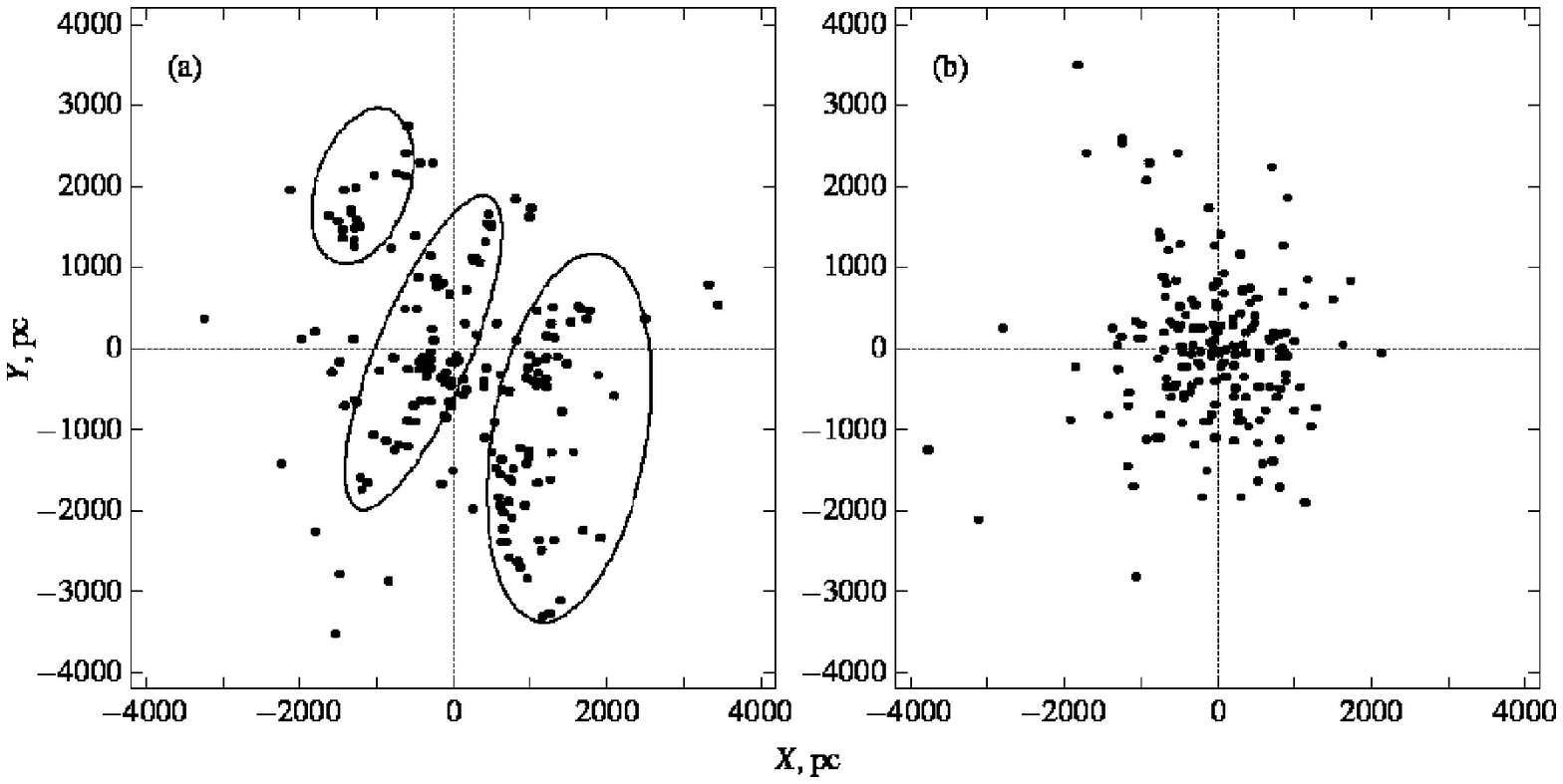}
\end{center}
} Fig. 1. Distribution of OSCs (a) younger than 50~Myr and (b)
older than 50~Myr in the Galactic $XY$ plane.
\end{figure}

\newpage
\begin{figure}[t]
{
\begin{center}
  \includegraphics[width=110mm]{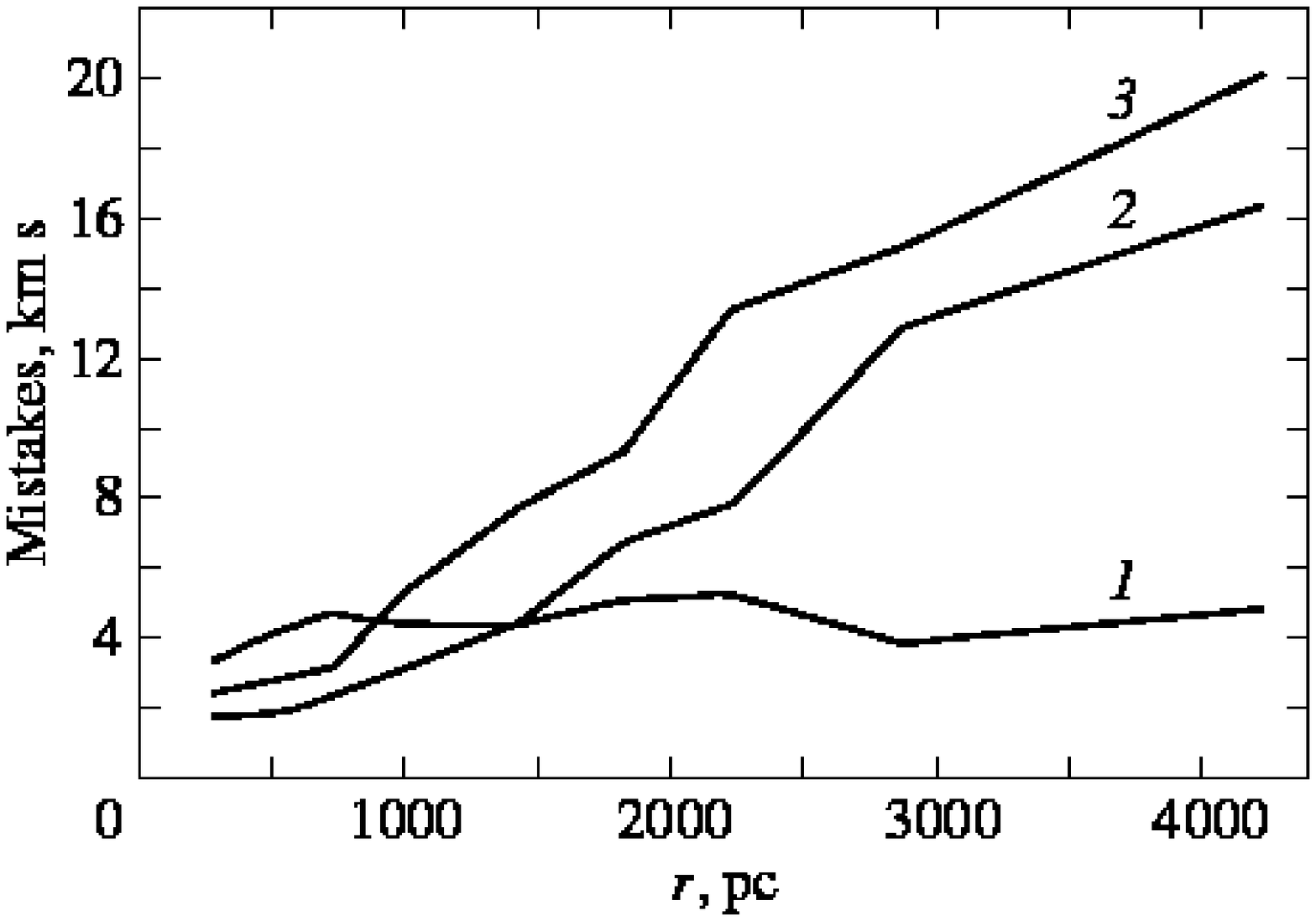}
\end{center}
} Fig. 2. Mean errors $\sigma_{V_r}$ (1), $\sigma_{V_l}$ (3), and
$\sigma_{V_b}$ (2) versus heliocentric distance.
\end{figure}
\begin{figure}[t]
{
\begin{center}
  \includegraphics[width=120mm]{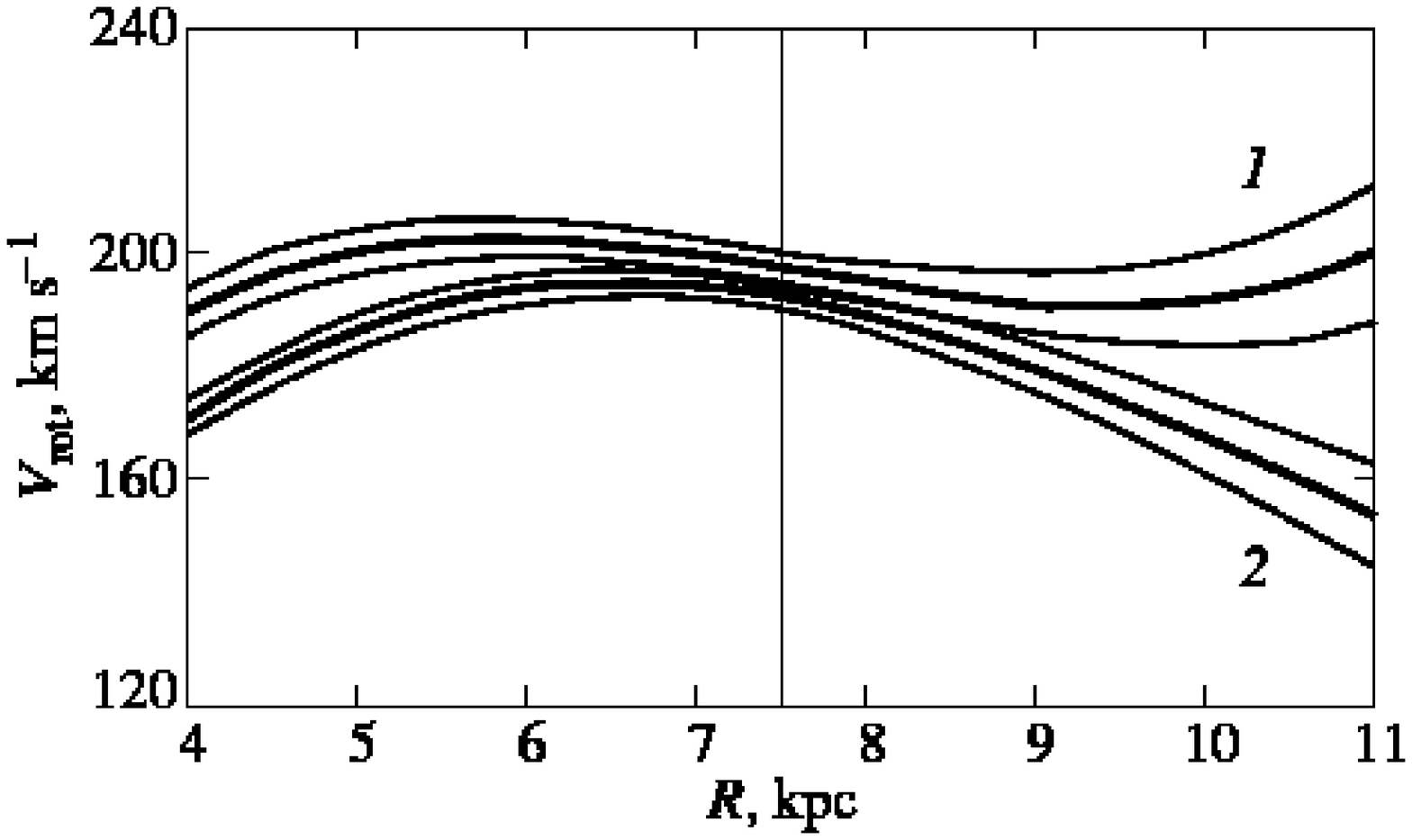}
\end{center}
} Fig. 3. Galactic rotation curve from clusters within 2.5 kpc (1)
and 5 kpc (2) of the Sun. The thin lines mark the boundaries of
the 1$\sigma$ confidence intervals; the vertical line indicates
the position of $R_0=7.4$ kpc.
\end{figure}

\newpage
\begin{figure}[t]
{
\begin{center}
  \includegraphics[width=160mm]{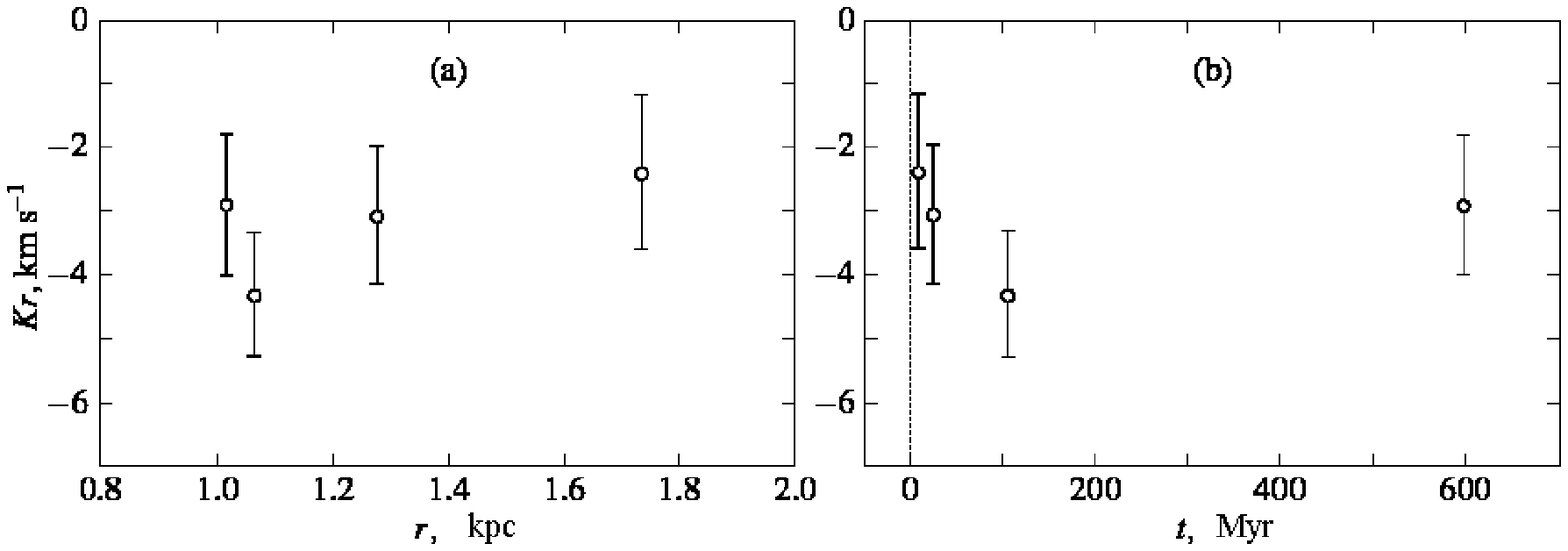}
\end{center}
} Fig. 4. Linear contraction velocity versus (a) heliocentric
distance and (b) mean sample age.
\end{figure}

\end{document}